\documentclass[aps,pra,twocolumn,amsfonts,amssymb,amsmath,showpacs,
floatfix,nofootinbib,groupedaddress,superscriptaddress,citesort]{revtex4}
\usepackage{mathrsfs}
\usepackage{amsfonts}
\usepackage{amstext}
\usepackage{amsmath}
\usepackage{amssymb}
\usepackage{bm}
\usepackage{bbm}
\usepackage[dvips]{graphicx}
\def\qed{\leavevmode\unskip\penalty9999 \hbox{}\nobreak\hfill
     \quad\hbox{\leavevmode  \hbox to.77778em{%
              \hfil\vrule   \vbox to.675em%
               {\hrule width.6em\vfil\hrule}\vrule\hfil}}
     \par\vskip3pt}

\begin{document}
\title
{Frozen condition of quantum coherence}

\author{Zhaofang Bai}
\email{baizhaofang@xmu.edu.cn} \affiliation{School of Mathematical
Sciences, Xiamen University, Xiamen, Fujian, 361000, China}

\author{Shuanping Du}\thanks{Corresponding author}
\email{dushuanping@xmu.edu.cn} \affiliation{School of Mathematical
Sciences, Xiamen University, Xiamen, Fujian, 361000, China}



\begin{abstract}

Quantum coherence as an important physical resource plays the key role in implementing various quantum tasks, whereas quantum coherence
 is often deteriorated due to the noise. In this paper, we analyse under which dynamical conditions the $l_1$-norm or the relative entropy of coherence can remain unchanged during the whole evolution (freezing coherence). 
  For single qubit systems, a nice formula is given to realize freezing coherence. Conversely, for a $d\  (d>2)$  dimensional system, we identify universal geometric conditions of freezing coherence.
 This offers an affirmative answer to the open question: how can one determine whether a unital quantum operation can be decomposed as a convex
combination of unitary operations [M. A. Nielsen and I. L. Chuang, Quantum Computation
and Quantum Information, (Cambridge University Press, Cambridge, 2000)]. Based on this analysis, we also give a complete classification of coherent states from operational coherence theory. This builds the counterpart of entanglement classification under LOCC.  
\end{abstract}


\maketitle

\section{Introduction }

Quantum coherence stands as one of the most essential features that
 results in nonclassical phenomena \cite{Leg1, Mandel}.
 It constitutes a powerful physical resource for implementing various tasks such as quantum algorithms \cite{Hillery,Matera,Pan,Wang,Ahnefeld,Karimi,Ye, Berberich,Escalera}, quantum metrology \cite{Giovannetti1,Giovannetti2,Demkowicz,Pires,
Cheng, Zhangt,Castellini,Ares, Lecamwasam}, quantum channel discrimination \cite{Girolami,Farace,Stre3,Takagi,Wilde,Rossi,Chen2}, witnessing quantum correlations \cite{ Mq,Hu1,Hu2,Mondal,Girolami2,Ding,Lee}, quantum phase transitions and  transport phenomena \cite{Karpat,Girolami3,Cakmak,Malvezzi,Chen3,Li2,Shi}.

A long-standing and significant issue concerning quantum coherence is the decoherence induced by the inevitable system-environment interaction. A great challenge in exploiting the resource is to fight against the decoherence since decoherence may weak the abilities of coherent states to fulfil quantum information processing tasks. Some propositions have been suggested for fighting against deterioration, such as  decoherence free subspaces \cite{Zanardi,Lidar}, quantum error correction codes \cite{Steane},  dynamical decoupling \cite{Viola},  quantum Zeno dynamics \cite{Rezakhani,Kwek}.
 Recently, Bromley et al. identify the conditions of sustaining long-lived quantum coherence \cite{Brom}, whose prediction has been confirmed in \cite{Silva,Zhangzhou}. It is  shown that
 the frozen condition for two qubits undergoing local identical
bit-flip operations with initial Bell-diagonal states is only dependent on the initial condition of the states \cite{Brom}.

 In a realistic physical system, atoms usually cannot be
dealt simply as noninteracting individual qubits when
atomic spacing is small. Thus, it is necessary to find the general frozen condition of quantum coherence \cite{Zhangzhou}.

In this note, we are aimed to study the frozen condition for $d$ dimensional coherent states under strongly incoherent operations (SIOs) \cite{Stre3,Benjamin,Winter}. The explicit geometric condition of freezing the $l_1$-norm or the relative entropy of coherence is revealed.
It shows that universal frozen condition is not only dependent on the arguments of coherent states but also dependent on the phases of
SIOs. The geometric condition is capable of selecting SIOs and coherent states with freezing coherence. Hence the ability of
such quantum states to perform quantum information processing tasks is not weakened under
the selected SIOs. In addition, we find that coherent states with the same argument can be frozen simultaneously. Based on the structural characterization of freezing the $l_1$-norm of coherence, we also provide the classification of coherent states which are in fact equivalent to freezing the relative entropy of coherence.

The paper is organized as follows. In section II, frozen condition is determined for coherent states with more nonzero off-diagonal elements.
In section III, we characterize frozen condition of X-states.  Section IV is a summary of our findings.

\section{Frozen condition of $\Omega$}

To present our finding clearly, we need first to recall the
basic formalism of the resource theory of coherence \cite{Stre3, Bau2}.
Free states are identified as
incoherent states which are defined as diagonal states in a prefixed basis  $\{|i\rangle\}_{i=1}^d$
for a $d$ dimensional Hilbert space ${\mathcal H}$, i.e., $$\rho=\sum_{i=1}^d\lambda_i|i\rangle\langle i|, \eqno{(1)}$$ $\lambda_i$
is a probability distribution.  The set of incoherent states will be
labelled by ${\mathcal I}$.
All other states which are not diagonal in
this basis are called coherent states.
Free operations are identified as incoherent operations which are
specified by a set of operation elements $\{K_l\}$ such that $K_l\rho K_l^\dag/Tr(K_l\rho K_l^\dag)\in {\mathcal
I}$ for all $\rho\in {\mathcal I}$, $$\Phi(\rho)=\sum_{l}K_l \rho K_l^\dag.\eqno{(2)}$$  Such operation elements $\{K_l\}$ are called incoherent. An incoherent operation  is a SIO if both $K_l$ and $K_l^\dag$ are incoherent \cite{Stre3,Benjamin,Winter}. We remark that there are different definitions of free operations stemming from meaningful physical and mathematical considerations \cite{Stre3}.
The category of  $\text{SIO}$ has appeared to be the most potential candidate of incoherent operations fulfilling desired norms of resource theory while in the meantime originated from physical motivation and experimentally enforceable.

For a $d\times d$ coherent state, one fundamental question is to ask the amount of
coherence it has. That is the quantification of coherence. Postulates for coherence quantifiers have been presented \cite{Stre3, Bau2}.
Any proper coherence measure $C$ is a non-negative function on quantum states and
must satisfy the following conditions:

$(C1)$ $C(\rho)\geq 0$ for every quantum state $\rho$, and $C(\rho)= 0$ for $\rho\in{\mathcal I}$;

$(C2a)$ Monotonicity under every $\text{ICO}$ $\Phi$:
$C(\Phi(\rho))\leq C(\rho)$,

or $(C2b)$ Monotonicity for average coherence under sub-selection
based on measurements  outcomes: $\sum_l p_l C(\rho_l)\leq C(\rho)$
for all $\{K_l\}$, where $\{K_l\}$ is from the operation representation of $\Phi$, $\rho_l=\frac{K_l\rho
K_l^\dag}{p_l}$ and $p_l={\rm Tr}(K_l\rho K_l^\dag)$.

$(C3)$ Convexity:
$C(\sum_l p_l\rho_l)\leq \sum_l p_l C(\rho_l)$ for any set of states
$\{\rho_l\}$ and any $p_l\geq 0$ with $\sum_l p_l=1$.\\

The coherence effect of a state is ascribed to the off-diagonal
elements of its density matrix with respect to the fixed basis $\{|i\rangle\}_{i=1}^d$.
Based on above mentioned postulates,
a very intuitive coherence measure
is the $l_1$-norm of coherence which is defined as
$C_{l_1}(\rho)=\sum_{i\neq j}|\rho_{ij}|, $
which is equal to a summation over the absolute values of all the off-diagonal elements
$\rho_{ij}$ of a given quantum state $\rho$ \cite{Bau2}.  Moreover, by virtue of the von Neumann entropy $S(\rho) = tr\rho \log\rho$, the relative
entropy of coherence is formed as $C_{Re}(\rho)=S(\rho)-S(\rho_{diag})$ \cite{Bau2}, where $\rho_{diag}$ denotes the diagonal state obtained from state $\rho$ by deleting all off-diagonal elements.

For a coherent state $\rho$ and a SIO $\Phi$, we say the $l_1$-norm of coherence is frozen if $C_{l_1}(\Phi(\rho))=C_{l_1}(\rho)$.
Similarly, if  $C_{Re}(\Phi(\rho))=C_{Re}(\rho)$, we say the relative entropy of coherence  is frozen. We are aimed to analyze under what dynamical conditions the coherence of quantum states can be frozen.

 Now, we  are ready to study  frozen behaviour of coherent states with more nonzero off-diagonal elements under SIOs.
 Typical example of such coherent states are superposed states  \cite{Ye,Bau2}, isotropic states \cite{Horodecki}, thermal states \cite{Weedbrook}.
 These states are widely used in quantum information processing


Denote  $\rho^{\sharp}=\{(i,j) \mid \rho_{ij}\neq 0,  i\neq j\}$ for any quantum state $\rho$ with $\rho_{ij}=\langle i| \rho|j\rangle$.
For $d=2$, let $$\Omega=\{\rho\mid (1,2)\in\rho^{\sharp}\}=\{\rho\mid\rho_{12}\neq 0\},\eqno{(3)}$$  the set of all coherent states. For $d\geq 3$, let $\Omega$ be the set of coherent states with
$$\left\{  \begin{array}{l}
 \cup_{(i,j)\in\rho^{\sharp}}\{i,j\}=\{1,2,\cdots, d\};\\
 \text{for } \forall\
(i,j)\in\rho^{\sharp},
\text{there  exists }\\
\quad (i,k)\text{\ or \ }(k,j)\in\rho^{\sharp}\ (k\neq i,j). \end{array} \right. \eqno{(4)} $$
It is easy to see that if $d=3$, then $$\Omega=\{\rho\mid \rho_{12}\rho_{13}\neq 0\}\cup \{\rho_{12}\rho_{23}\neq 0\}\cup \{\rho_{13}\rho_{23}\neq 0\}.$$
 Note that for a SIO,  every operation element $K_l$ can be written as the form
$$K_{l}=\sum_{i=1}^ d d_{l i}|f_{l}(i)\rangle \langle i|,\eqno{(5)}$$ $f_{l}$ being permutations of $\{1,2,\ldots, d\}$.

Now, we are in a position to give our main result.

\vspace{0.1in}
{\bf Theorem 2.1.} {\it For $\rho\in \Omega$ and a SIO $\Phi$ with $\Phi(\rho)\in \Omega$, then
 $C_{l_1}(\rho)=C_{l_1}(\Phi(\rho))$ if and only if $${\Phi(\cdot})=\sum_{l} \delta_{l} U_{l}\cdot U^{\dag}_{l}, U_{l}=\sum_{i=1}^ d e^{{\rm i}\theta_i^{(l)}}|f_{l}(i)\rangle \langle i|, \eqno{(6)}$$ $\delta_{l}>0, \sum_l \delta_{l}=1 $. In addition, for $\forall$  $ 1 \leq m\neq n\leq d$,
$$ \Big|\sum_{\substack{
l,i,j\\
f_{l}(i)=m,\  f_{l}(j)=n}} e^{{\rm i}(\theta_i^{(l)}-\theta_j^{(l)})}\rho_{ij}\Big|=\sum_{\substack {
l,i,j\\
f_{l}(i)=m,\ f_{l}(j)=n}}
|\rho_{ij}|. \eqno{(7)}$$ }
\vspace{0.1in}

In quantum information theory, an open question is, given a unital quantum operation, how can
one determine whether this operation can be decomposed as a convex combination of unitary
operations \cite[Page 396, Problem 8.3]{Nielc}. The physical motivation behind this key question is the desire to distinguish various error mechanisms afflicting the preparation and processing of quantum states. 
Despite considerable effort (see \cite{Mendlx} and references therein), not much is known about the explicit structure of convex combination of unitary
operations beyond single qubit systems (in which case all unital operations are convex combination of unitary
operations).
For larger systems, this is no longer true \cite{Mendlx}. By (6) of Theorem 2.1,  any SIO  freezing the $l_1$-norm of coherence of some coherent state in $\Omega$ has  desired decomposition.  This shows that the only error mechanism occurring in the realization of coherent states of $\Omega$ is a classical uncertainty, thus the resulting gate is described by the convex combination of unitary operations.


On the other hand,
for arbitrary $1\leq m\neq n\leq d$, (7) shows that the triangle inequality for
complex numbers  $$\{e^{{\rm i}(\theta_i ^{(l)}-\theta_j^{(l)})}\rho_{ij}| f_l(i)=m, f_l(j)=n, l=1,2,\cdots\} \eqno (8)$$ is in fact a equality. This implies that such complex numbers have equal argument which reveals geometric feature of freezing the $l_1$-norm of coherence. The geometric feature is efficient at selecting SIOs and coherent states with freezing coherence. In addition,
an interesting observation of (7) shows any coherent state having the same argument of corresponding entries with $\rho$ can be frozen also by $\Phi$.


Let us consider freezing the $l_1$-norm of coherence for one-qubit systems.
For a coherent state  $\rho=(\rho_{ij}) (1\leq i,j\leq 2)$ and  a SIO $\Phi$, by  Theorem 2.1, we have $$\begin{array}{ll}
C_{l_1}(\rho)=C_{l_1}(\Phi(\rho))\Leftrightarrow & {\Phi(\rho})= \delta U_{1}\rho U^{\dag}_{1} +(1-\delta )U_{2}\rho U^{\dag}_{2},\\ &\theta_1+\theta_2+2\theta=2k\pi,\end{array}\eqno (9)$$  here $U_{1}=\left(\begin{array}{cc} e^{i\theta_1} &0 \\0& 1\end{array}\right) $, $U_{2}=\left(\begin{array}{cc} 0& 1\\e^{i\theta_2}&0\end{array}\right)$, $\theta$ is the the argument of nonzero $\rho_{12}$, $k$ is some integer.
In quantum information processing, researchers are
usually interested only in some paradigmatic SIOs,  such as bit-flip operations,  bit+phase-flip operations, phase-flip operations and depolarizing operations. Conditions of $(9)$ imply that paradigmatic bit-flip operations $(\theta_1=\theta_2=0)$ and bit+phase-flip operations $(\theta_1=0, \theta_2=\pi)$ allow  freezing coherence for specific coherent states, while phase-flip operations and depolarizing operations can not realize freezing coherence. Frozen conditions for  coherent states
under bit-flip operations and bit+phase-flip operations have been provided in \cite{Brom}. These conditions are expressed in terms of Bloch vectors of coherent states. A direct computation shows that such frozen conditions are equivalent to $\theta_1+\theta_2+2\theta=2k\pi$. Compared with the particular SIOs \cite{Brom}, (9) offers all possible SIOs for freezing coherence.

Next, we consider frozen condition of isotropic states on two-qubit systems \cite{Terhal,Rungta,Du2}. Assume $$\rho=\frac{1-\alpha}{3}I+\frac{4\alpha-1}{3}|\psi_+\rangle\langle\psi_+|, \alpha\in (\frac{1}{4},1],\eqno{(10)}$$ $|\psi_{+}\rangle=(\frac{1}{2}, \frac{1}{2}, \frac{1}{2}, \frac{1}{2})^t,$  $$\Phi(\rho)=\delta U_1\rho U_1^\dag +(1-\delta)U_2\rho U_2^\dag, 0<\delta<1\eqno{(11)}$$
$$U_1=e^{i\theta_1}|2\rangle\langle 1|+e^{i\theta_2}|1\rangle\langle 2|+e^{i\theta_3}|3\rangle\langle 3|+e^{i\theta_4}|4\rangle\langle 4|$$
$$U_2=e^{i\theta_5}|3\rangle\langle 1|+e^{i\theta_6}|4\rangle\langle 2|+e^{i\theta_7}|1\rangle\langle 3|+e^{i\theta_8}|2\rangle\langle 4|.$$
By Theorem 2.1, one have $$C_{l_1}(\rho)=C_{l_1}(\Phi(\rho))\Leftrightarrow\left\{\begin{array}{l}
\theta_2-\theta_1+\theta_7-\theta_8=2k_1\pi\\
\theta_2-\theta_3+\theta_7-\theta_5=2k_2\pi\\
\theta_2-\theta_4+\theta_7-\theta_6=2k_3\pi\\
\theta_1-\theta_3+\theta_8-\theta_5=2k_4\pi\end{array},\right.\eqno{(12)}$$
$k_i(i=1,2,3,4)$ are some integers. In fact, if the $l_1-$norm of coherence of isotropic states are frozen, then the $l_1-$norm of coherence of any coherent state with positive entries is frozen too.

It is worth noting that Theorem 2.1 is applicable
to all local operations whose operation elements consisting of the typical qubit noisy operations including the bit-flip, phase-flip, bit+phase-flip, depolarizing, phase damping, and amplitude damping operations \cite{Xiao}. Recall that  a local bit-flip operation of an $N$-qubit system has the form
$$\Phi_t=\Phi_{q_1}^1\otimes \cdots \otimes \Phi_{q_N}^N,\eqno(13)$$ where $\Phi_{q_l}^l(\rho)=(1-q_l)\rho+q_l\sigma_1\rho\sigma_1$, and $q_1, \cdots , q_N$ are parameters dependent on time,
 $\sigma_1=\left(\begin{array}{cc}
                                                                                                   0 & 1\\
                                                                                                   1& 0\end{array}\right)$ is the Pauli-X operator.
It is easy to check each local bit-flip operations is a SIO. Therefore we can apply Theorem 2.1 to local bit-flip operations.
Let us  consider a 2-qubit system undergoing a local bit flip operation  $\Phi$
which has operation elements  $$\begin{array}{ll} K_1=(1-\frac q2)I_2\otimes I_2, & K_2=\sqrt{\frac q2(1-\frac q2)}I_2\otimes \sigma_1,\\  K_3=\sqrt{\frac q2(1-\frac q2)}\sigma_1\otimes I_2, & K_4=\frac q2 \sigma_1\otimes \sigma_1.\end{array} \eqno(14)$$
Theorem 2.1 can help us to effectively identify coherent states
with frozen the $l_1$-norm of coherence. Assume $\rho=(\rho_{ij})\in \Omega, \Phi(\rho)\in\Omega$,  by a direct computation, one can see that $$\begin{array}{ll}C_{l_1}(\Phi(\rho))=C_{l_1}(\rho)\Leftrightarrow &\rho_{ij}\in{\mathbb R}(1\leq i,j\leq4), \rho_{12}\rho_{34}\geq 0,\\
 & \rho_{13}\rho_{24}\geq 0, \rho_{14}\rho_{23}\geq 0.\end{array}\eqno(15)$$

%




\vspace{0.1in}

 In the following, we are aimed to characterize dynamical conditions of freezing the relative entropy of coherence.
 It is found freezing the relative entropy of coherence is equivalent to the classification of coherent states.

 Before giving our result, let us recall briefly the classification of coherent states.
 In view of resource theory \cite{Stre3,Bennett,Vidal,Wolf,Du}, one
fundamental issue is the classification of coherent
states. A natural way of defining equivalence relations in
the set of coherent states is that equivalent states
contain the same amount of coherence.  The monotonicity
of coherence under SIOs allows
us to identify any two states that can be transformed from
each other  by SIOs.
Clearly, this criterion is interesting in quantum information
theory, since equivalent states are indistinguishable for exactly
the same tasks.

{\bf Definition 2.2.} For coherent  states $\rho, \sigma\in\Omega$, we say
they are SIO equivalent if there exist SIOs $\Phi$ and  $\Psi$ satisfying
$\Phi(\rho)=\sigma$ and $\Psi(\sigma)=\rho$. We denote this relation by $\rho\stackrel{\text {SIO}}{\sim}\sigma$.

Based on Theorem 2.1, we can provide a complete
classification  of coherent states in $\Omega$.

\vspace{0.1in}
{\bf Theorem 2.3.} {\it Assume  $\rho,\sigma\in\Omega$, then the
following statements are equivalent:

(i) $\rho\stackrel{\text {SIO}}{\sim}\sigma$;

(ii) There exists a strictly incoherent  unitary operation $U=P_\pi\left(\begin{array}{ccc}
                           e^{i\theta_1} & & 0\\
                               &\ddots &\\
                               0& & e^{i\theta_d}\end{array}\right)$ such that $\Phi(\rho)=U\rho U^\dag$, here $P_{\pi}$ is the permutation matrix corresponding to a permutation $\pi$ of $\{1,2,\cdots, d\}$, $\theta_i\in{\mathbb R}, 1\leq i\leq d$;

(iii) There is a SIO $\Phi$ satisfying $\Phi(\rho)=\sigma$ and $C_{RE}(\rho)=C_{RE}(\sigma)$.}
\vspace{0.1in}

By (i) and (ii) of Theorem 2.3, two coherent states are equivalent if and only if they are related by strictly incoherent unitary operations.
A parallel and beautiful result in quantum entanglement is that two pure entangled
states are equivalent (they can be transformed from each other by LOCC) if
and only if they are related by local unitary operations \cite{Bennett,Vidal}.
For $\rho, \Phi(\rho)\in\Omega$, (ii) and (iii) of Theorem 2.3 show that strictly incoherent unitary
operation is a necessary consequence of freezing the relative entropy of coherence. This implies that
if the relative entropy of coherence is frozen, then all measures of coherence are
frozen \cite{Xiao}. We note that frozen coherence is dependent on the coherence measures adopted in general \cite{Brom, Xiao}
 Some coherence measures being frozen do not
imply other coherence measures being frozen too, since
different coherence measures result in different orderings of
coherence \cite{Tong2}.

\section{Frozen condition  of X-states}

Theorem 2.1 and Theorem 2.3 identify frozen condition of coherent states with more nonzero off-diagonal elements. An
interesting question is to find frozen condition
of coherent states with less nonzero off-diagonal elements. The typical example of
such states are X-states which are frequently used as important resources for the realization of various tasks related to quantum communication and computation \cite{Nielc,Haddadi,Balthazar}. Let $$\Omega_\text{X}=\{\rho\ | \rho_{i,d+1-i}\neq0, \rho_{ii}\neq 0, i=1,2,\cdots, d \},\eqno(16)$$ we identify frozen condition of $\Omega_\text{X}$ in the section. 

A key tool for characterizing freezing coherence on $\Omega_\text {X}$ is the direct sum decomposition of X-states.
Recall that if $d $ is even, then each X-state can be decomposed as the direct sum of $d/2$ quantum
states
$$P_{\pi}\rho P_{\pi}^{\dag}=\lambda_1\rho_1\oplus \lambda_2\rho_2 \oplus\cdots  \oplus\lambda_{\frac d 2}\rho_{\frac d 2} \eqno(17)$$ up to some permutation $P_{\pi}$, with $$\rho_i =\frac 1{\rho_{ii}+\rho_{(d+1-i)(d+1-i)}}\left( \begin{array}{cc} \rho_{ii} & \rho_{i(d+1-i)}\\ \rho_{(d+1-i)i}& \rho_{(d+1-i)(d+1-i)}\end{array}\right),$$ $$\lambda_i=\rho_{ii}+\rho_{(d+1-i)(d+1-i)}$$ and
$$\pi(i)=2i-1, \pi(d+1-i)=2i,\ i=1,2,\ldots,\frac d 2.$$
If $d$ is odd, every X-state can be
decomposed as the direct sum of $[\frac d 2]$ quantum states plus an additional one dimensional
matrix $$P_{\pi}\rho P_{\pi}^{\dag}=\lambda_1\rho_1\oplus \lambda_2\rho_2 \oplus\cdots  \oplus\lambda_{[\frac d 2]}\rho_{[\frac d 2]}\oplus \lambda_{[\frac d 2]+1}\eqno(18)$$ up to some permutation $P_{\pi}$, where $[\frac d 2]$ denotes the integer part of the number $\frac d 2$ and $$\pi(i)=2i-1,\pi(d+1-i)=2i, \pi([\frac d 2]+1)=d,\  i=1,\ldots,[\frac d 2].$$

\vspace{0.1in} To better understand how our general result works, we will give an example for the two-qubit systems.
It is easy to see  $$P_{\pi}=\left(\begin{array}{cccc}1& 0&0&0\\0&0&0&1\\0&1&0&0\\0&0&1&0\end{array}\right),$$
$$\begin{array}{l}P_{\pi}\left(\begin{array}{cccc}\rho_{11}& 0&0&\rho_{14}\\0&\rho_{22}&\rho_{23}&0\\0&\rho_{32}&\rho_{33}&0\\\rho_{41}&0&0&\rho_{44}\end{array}\right)P_{\pi}^\dag
=\left(\begin{array}{cccc}\rho_{11}& \rho_{14}&0&0\\\rho_{41}&\rho_{44}&0&0\\0&0&\rho_{22}&\rho_{23}\\0&0&\rho_{32}&\rho_{33}\end{array}\right)
\\=\frac{1}{\rho_{11}+\rho_{44}}\rho_1+\frac{1}{\rho_{22}+\rho_{33}}\rho_2.\end{array}$$
It is evident that $$\begin{array}{cc}\rho\mapsto \Phi(\rho)=\sum _l K_l\rho K_l^\dag &\\
\Updownarrow & \\
 P_{\pi}\rho P_{\pi}^\dag\mapsto\sum _l P_{\pi}K_lP_{\pi}^\dag P_{\pi}\rho P_{\pi}^\dag P_{\pi}K_l^\dag P_{\pi}^\dag.\end{array}\eqno{(19)}$$
Therefore we may assume that $\rho$ is the direct sum of two quantum states and each Kraus operator of $\Phi$ has the form $P_{\pi}K_lP_{\pi}^\dag$.
Let $$P_{\pi}K_1P_{\pi}^\dag=\left(\begin{array}{cccc}0& \delta_1e^{i\theta_2}&0&0\\ \delta_1e^{i\theta_1}&0&0&0\\0&0&\delta_2e^{i\theta_3}&0\\0&0&0&\delta_2e^{i\theta_4}\end{array}\right),$$ $$P_{\pi}K_2P_{\pi}^\dag=\left(\begin{array}{cccc}0& 0&\delta_4e^{i\theta_7}&0\\ 0&0&0&\delta_4e^{i\theta_8}\\ \delta_3e^{i\theta_5}&0&0&0\\0&\delta_3e^{i\theta_6}&0&0\end{array}\right),$$ $\delta_1>0,$ $\delta_2>0, \delta_1^2+\delta_3^2=\delta_2^2+\delta_4^2=1$. By a direct computation, we can get
$$\begin{array}{c}C_{l_1}(\rho)= C_{l_1}(\Phi(\rho))\\
\Updownarrow\\
\left\{\begin{array}{l}
|e^{i(\theta_2-\theta_1)}\overline{\rho_{14}}+e^{i(\theta_7-\theta_8)}\rho_{23}|=|\rho_{14}|+|\rho_{23}|\\
|e^{i(\theta_3-\theta_4)}\rho_{23}+e^{i(\theta_5-\theta_6)}\rho_{14}|=|\rho_{23}|+|\rho_{14}|.
\end{array}\right.\end{array}\eqno{(20)}$$
Denote $$U_1=\left(\begin{array}{cc} e^{i\theta_1}&0\\0&e^{i\theta_2}\end{array}\right), U_2=\left(\begin{array}{cc} e^{i\theta_3}&0\\0&e^{i\theta_4}\end{array}\right),$$ $$U_3=\left(\begin{array}{cc} e^{i\theta_5}&0\\0&e^{i\theta_6}\end{array}\right),U_4=\left(\begin{array}{cc} e^{i\theta_7}&0\\0&e^{i\theta_8}\end{array}\right).$$
Combining the additivity of relative entropy of
coherence for subspace independent states \cite{Yuzhang} with the equality for convexity of relative entropy \cite{Jencova}, we have
$$C_{Re}(\rho)= C_{Re}(\Phi(\rho))\Leftrightarrow\left\{\begin{array}{l}
U_1\rho_1 U_1^\dag=U_4\rho_2 U_4^\dag\\
U_2\rho_2 U_2^\dag=U_3\rho_1 U_3^\dag.
\end{array}\right.\eqno{(21)}$$
It is amazing that frozen conditions (20), (21) and the forms of Kraus operators are indicative of our general results.


\vspace{0.1in}
{\bf Theorem 3.1.} {\it If $d$ is even, $\rho,\Phi(\rho)\in \Omega_\text{X}$ and strictly incoherent operation $\Phi$  with the Kraus form ${\Phi(\rho})= \sum_{l}K_{l}\rho K_{l}^\dag$,  then $C_{l_1}(\rho)=C_{l_1}(\Phi(\rho))$ if and only if
$$P_{\pi}K_{l}P_{\pi}^{\dag}= P_{f_{l}}\cdot \oplus_{i=1}^{\frac d 2} {\delta}_{l i}U_{l i},\eqno{(22)}$$ where $P_{f_{l}}$ is the
permutation matrix corresponding to some permutation $f_{l}$ of $\{1,2,\ldots,d\}$,
 ${\delta}_{l i}\geq 0$, $U_{l i}=\left(\begin{array}{cc} e^{i\theta_{l i1}}& 0\\0& e^{i\theta_{l i2}}\end{array}\right)$, $\sum_l\delta_{l i}^2=1$ for every $1\leq i\leq\frac{d}{2}$. In addition, for $\forall i$, there is $1\leq m\leq \frac{d}{2}$ such that $f_{l}(\{2i-1,2i\})=\{2m-1,2m\}$,
and for given $ m$,  $$ \begin{array}{l}\Big|\sum_{l_{f_{l}(\{2i-1,2i\})=\{2m-1,2m\}}}\delta_{l i}^2e^{i(\theta_{l i1}-\theta_{l i2})}\rho_{2i-1\ 2i}\Big|\\=\sum_{l_{f_{l}(\{2i-1,\ 2i\})=\{2m-1,\ 2m\}}}\delta_{l i}^2|\rho_{2i-1\ 2i}|.\end{array}\eqno{(23)}$$}


{\bf Theorem 3.2.} {\it If $d$ is even, $\rho,\sigma\in \Omega_\text{X}$ and strictly incoherent operation $\Phi$  with the Kraus form ${\Phi(\rho})= \sum_{l}K_{l}\rho K_{l}^\dag$,  then $C_{Re}(\rho)=C_{Re}(\Phi(\rho))$ if and only if
$$P_{\pi}K_{l}P_{\pi}^{\dag}= P_{f_{l}}\cdot \oplus_{i=1}^{\frac d 2} {\delta}_{l i}U_{l i},\eqno{(24)}$$  where $P_{f_{l}}$ is the
permutation matrix corresponding to some permutation $f_{l}$ of $\{1,2,\ldots,d\}$,
 ${\delta}_{l i}\geq 0$, $U_{l i}=\left(\begin{array}{cc} e^{i\theta_{l i1}}& 0\\0& e^{i\theta_{l i2}}\end{array}\right)$, $\sum_l\delta_{l i}^2=1$ for every $1\leq i\leq\frac{d}{2}$. In addition, for $\forall i$, there is $m$ such that $f_{l}(\{2i-1,2i\})=\{2m-1,2m\}$,
and for given $m$, $$U_{l i}\rho_i U_{l i}^{\dag}=U_{s j}\rho_j U_{s j}^{\dag}\eqno{(25)}$$ for $i,j$ with $f_{l}(\{2i-1,2i\})=f_{s}(\{2j-1,2j\})=\{2m-1,2m\}$.}

{\bf Theorem 3.3.} {\it If $d$ is odd, $\rho,\Phi(\rho)\in \Omega_\text{X}$ and strictly incoherent operation $\Phi$  with the Kraus form ${\Phi(\rho})= \sum_{l}K_{l}\rho K_{l}^\dag$,    then $C_{l_1}(\rho)=C_{l_1}(\Phi(\rho))$ if and only if
$$P_{\pi}K_{l}P_{\pi}^{\dag}=P_{f_{l}}\cdot \oplus_{i=1}^{[\frac d 2]} {\delta}_{l i}U_{l i}+\delta_{l,d}|d\rangle\langle d|\eqno{(26)}$$ or $$P_{\pi}K_{l}P_{\pi}^{\dag}=\delta_{l,d}|i\rangle\langle d|\eqno{(27)}$$ for some $1\leq i< d$, here $U_{l i}$, $f_{l}$ and $\delta_{l i}$  have the same property as that of Theorem 3.1. }

{\bf Theorem 3.4.} {\it If $d$ is odd, $\rho,\Phi(\rho)\in \Omega_\text{X}$ and strictly incoherent operation $\Phi$  with the Kraus form ${\Phi(\rho})= \sum_{l}K_{l}\rho K_{l}^\dag$,  then $C_{Re}(\rho)=C_{Re}(\Phi(\rho))$ if and only if
$$P_{\pi}K_{l}P_{\pi}^{\dag}=P_{f_{l}}\cdot \oplus_{i=1}^{[\frac d 2]} {\delta}_{l i}U_{l i}+\delta_{l,d}|d\rangle\langle d|\eqno{(28)}$$ or $$P_{\pi}K_{l}P_{\pi}^{\dag}=\delta_{l,d}|i\rangle\langle d|\eqno{(29)}$$ for some $1\leq i< d$, here $U_{l i}$, $f_{l}$ and $\delta_{l i}$ have the same property as that of Theorem 3.2.}

In order to understand Theorem 3.3 and Theorem 3.4, let us consider the case $d=3$.  From Theorem 3.3,  every Kraus operator $P_{\pi}K_lP_{\pi}^\dag$ must have one of the following forms:
$$\begin{array}{ll}
(1)\left(\begin{array}{ccc} \delta_{11}e^{i\theta_1} &0 &0\\0& \delta_{11}&0\\0&0& \delta_{13}\end{array}\right) , &(2) \left(\begin{array}{ccc} 0& \delta_{21}&0\\ \delta_{21}e^{i\theta_2}&0&0\\0&0&\delta_{23}\end{array}\right), \\
(3) \left(\begin{array}{ccc} 0 &0 &\delta_{33}\\0& 0&0\\0&0& 0\end{array}\right) , &(4) \left(\begin{array}{ccc} 0 &0 &0\\0& 0&\delta_{43}\\0&0& 0\end{array}\right).\end{array}$$ A direct computation shows
$$\begin{array}{c}C_{l_1}(\rho)=C_{l_1}(\Phi(\rho))\\
\Updownarrow \\
\theta_1+\theta_2+2\theta=2k\pi,k\in{\mathcal Z}, \rho_{13}=|\rho_{13}|e^{i\theta}.\end{array}\eqno{(30)}$$
Using Theorem 3.4, we can obtain $$\begin{array}{c}C_{Re}(\rho)=C_{Re}(\Phi(\rho))\\
\Updownarrow \\
\rho_{11}=\rho_{22}, \theta_1+\theta_2+2\theta=2k\pi,k\in{\mathcal Z}, \rho_{13}=|\rho_{13}|e^{i\theta}.\end{array}\eqno{(31)}$$

Applying Theorem 3.2, we can get an important result of \cite{Brom} which considers freezing the relative entropy of coherence for $2$-qubit systems under local bit-flip operations.
Borrow the notations in \cite{Brom}, if the initial coherent state $\rho$ is Bell-diagonal \cite{RMHorodecki}, then
$$\rho=\frac 1 4(I\otimes I+\sum_{i=1}^3c_j\sigma_j\otimes \sigma_j),\eqno{(32)}$$here $\sigma_j$ is the Pauli matrix. Then
$$\rho=\frac 14 \left(\begin{array}{cccc} 1+c_3 & 0 &0&c_1-c_2\\
                                         0&1-c_3 & c_1+c_2&0 \\
                                         0& c_1+c_2 &1-c_3& 0 \\
                                         c_1-c_2 & 0 &0&  1+c_3 \end{array}\right).$$
The local bit-flip operation has the Kraus operators $$\begin{array}{ll} K_1=(1-\frac q2)I_4, & K_2=\sqrt{\frac q2(1-\frac q2)}I_2\otimes \sigma_1,\\  K_3=\sqrt{\frac q2(1-\frac q2)}\sigma_1\otimes I_2, & K_4=\frac q2 \sigma_1\otimes \sigma_1.\end{array}$$
 A direct computaion shows that
 $$\begin{array}{l} P_{\pi}K_1P_{\pi}^{\dag}=(1-\frac q2)I_4=(1-\frac q2)\left(\begin{array}{cc} I_2&0\\0&I_2\end{array}\right), \\ P_{\pi}K_2P_{\pi}^{\dag}=\sqrt{\frac q2(1-\frac q2)}\left(\begin{array}{cc} 0&I_2\\ I_2 &0 \end{array}\right),\\
 P_{\pi} K_3P_{\pi}^{\dag}=\sqrt{\frac q2(1-\frac q2)}\left(\begin{array}{cc} 0&\sigma_1\\ \sigma_1&0\end{array}\right),\\
 P_{\pi} K_4 P_{\pi}^{\dag}=\frac q2 \left(\begin{array}{cc} \sigma_1 &0 \\ 0&\sigma_1\end{array}\right),\end{array}$$ From  Theorom 3.2, it follows that
 $$C_{Re}(\rho)=C_{Re}(\Phi(\rho))\Leftrightarrow\frac{c_1+c_2}{2-2c_3}=\frac{c_1-c_2}{2+2c_3}.\eqno{(33)}$$ That is $c_2=-c_1c_3$ which is an important conclusion of \cite{Brom}.
Compared to the scheme of choosing particular SIOs \cite{Brom}, we can provide all possible SIOs for realizing frozen coherence.

\vspace{0.1in} \section{Summary}

Complex systems are inevitably subject to noise, hence
it is natural and technologically crucial to question under
what conditions quantum coherence is not deteriorated during open evolutions \cite{Buchleitner}.
In this work, we have determined exact conditions such that the $l_1$-norm or the relative entropy of coherence
is dynamically frozen \cite{Brom,Zhangzhou}. It shows that frozen conditions are dependent on the arguments
of coherent states and the phases of SIOs. This give us the ability to
identify coherent states and SIOs with frozen behaviour.
It has been shown that
a bigger amount of coherence decreases the failure of Deutsch-Jozsa algorithm \cite{Hillery,Pan}.
Necessary and sufficient conditions for coherent states to be equivalent
under SIOs are derived. It is shown that two coherent states are equivalent under SIOs if and only if they
are related by strictly incoherent unitary operations. This builds the counterpart of the celebrated result that two
pure entangled states are equivalent under local operations and classical communication (LOCC) if and only if
they are related by local unitaries \cite{Vidal}.

\vspace{0.1in}{\bf Acknowledgements}

\vspace{0.1in} This research was supported by NSF of China (12271452,11671332), NSF of
Fujian (2023J01028) and NSF of Xiamen (3502Z202373018).

\vspace{0.1in}

{\it\bf Appendix: Proof of main results.}\vspace{0.1in}

Proofs of all results in this paper are given in appendix.

{\bf Proof of Theorem 2.1.}
``$\Leftarrow $'' Assume $\Phi(\rho)=\sigma$, by a direct computation of $\Phi(\rho)$, one can see the entries of $\sigma$ have the following form: $$\sigma_{mn}=\Phi(\rho)_{mn}=\sum_{\substack{
l,i,j\\
f_{l}(i)=m,f_{l}(j)=n}}\delta_l e^{{\rm i}(\theta_i^{(l)}-\theta_j^{(l)})}\rho_{ij}.$$
From  Eq. (7), one have
$$\begin{array}{ll}
     & C_{l_1}(\sigma)=\sum_{m\neq n}|\sigma_{mn}|\\
=    & \sum_{m\neq n}\Big|\sum_{i\neq j}\sum_{l_{f_{l}(i)=m,f_{l}(j)=n}}\delta_l e^{{\rm i}(\theta_i^{(l)}-\theta_j^{(l)})}\rho_{ij}\Big|\\
=& \sum_{m\neq n}\sum_{i\neq j}\sum_{l_{f_{l}(i)=m,f_{l}(j)=n}}\delta_{l}|\rho_{ij}|\\
=& \sum_{l}\sum_{i\neq j,\ m\neq n, \ f_{l}(i)=m,f_{l}(j)=n}\delta_{l}|\rho_{ij}|\\
=&  \sum_{l}\sum_{ m\neq n}\delta_{l}|\rho_{f_{l}^{-1}(m)f_{l}^{-1}(n)}|\\
=& \sum_{l}\sum_{i\neq j}\delta_{l}|\rho_{ij}|=C_{l_1}(\rho).
\end{array}$$

``$\Rightarrow $'' Now we assume $\Phi$ have the Kraus operators $K_{l}=\sum_{i=1}^ d d_{l,i}|f_{l}(i)\rangle \langle i|$.
 A direct computation shows that the following formulas hold true: $$\begin{array}{ll}
     & C_{l_1}(\sigma)=\sum_{m\neq n,\ \sigma_{mn}\neq 0}|\sigma_{mn}|\\
=    & \sum_{m\neq n,\ \sigma_{mn}\neq 0}\Big|\sum_{i\neq j}\sum_{l_{f_{l}(i)=m,f_{l}(j)=n}}d_{l,i}\overline {d_{l,j}}\rho_{ij}\Big|\\
\overset{(34)}\leq  & \sum_{m\neq n,\sigma_{mn}\neq 0}\sum_{i\neq j}(\sum_{l_{f_{l}(i)=m,f_{l}(j)=n}}|d_{l,i}\overline {d_{l,j}}|)|\rho_{ij}|\\
=    &    \sum_{i\neq j}\sum_{m\neq n,\sigma_{mn}\neq 0}(\sum_{l_{f_{l}(i)=m,f_{l}(j)=n}}|d_{l,i}\overline {d_{l,j}}|)|\rho_{ij}|\\
=   &  \sum_{i\neq j}(\sum_{m\neq n,\sigma_{mn}\neq 0}\sum_{l_{f_{l}(i)=m,f_{l}(j)=n}}|d_{l,i}\overline {d_{l,j}}|)|\rho_{ij}|\\
\overset{(35)}\leq & \sum_{i\neq j}(\sum_{m\neq n}\sum_{l_{f_{l}(i)=m,f_{l}(j)=n}}|d_{l,i}\overline {d_{l,j}}|)|\rho_{ij}|\\
=    &  \sum_{i\neq j}(\sum_{l}|d_{l,i}\overline {d_{l,j}}|)|\rho_{ij}|\\
\overset{(36)}\leq  &  \sum_{i\neq j,\ \rho_{ij}\neq 0}\sqrt{\sum_{l}|d_{l,i}|^2\sum_{l}|d_{l,j}|^2}|\rho_{ij}|\\
=    & \sum_{i\neq j} |\rho_{ij}|=C_{l_1}(\rho). \end{array}$$
By $C_{l_1}(\rho)=C_{l_1}(\sigma)$, it follows that the inequalities $(34)$, $(35)$ and $(36)$, in fact, are equalities.
For $\rho_{ij}\neq 0$, combining the Cauchy-Schwarz inequality and $(36)$, we obtain   $|d_{l,i}|=|d_{l,j}|$ for arbitrary $l$ and $1\leq i,j\leq d$. By the property of $\Omega$, we have  $|d_{l,i}|=|d_{l,j}|$ for arbitrary $l$ and $1\leq i,j\leq d$.
Denote $|d_{l,i}|^2=\delta_{l}$,  $d_{l,i}=\sqrt{\delta_{l}}e^{{\rm i}\theta_i^{(l)}},$ and  $U_{l}=\sum_{i=1}^ d e^{{\rm i}\theta_i^{(l)}}|f_{l}(i)\rangle \langle i|$, then  $K_{l}=\sqrt{\delta_{l}} U_{l}$
is the desired. The equality $$ \Big|\sum_{\substack{
l,i,j\\
f_{l}(i)=m,f_{l}(j)=n}}\delta_l e^{{\rm i}(\theta_i^{(l)}-\theta_j^{(l)})}\rho_{ij}\Big|=\sum_{\substack {
l,i,j\\
f_{l}(i)=m,f_{l}(j)=n}}
\delta_{l}|\rho_{ij}|.$$ follows from $(34)$,  i.e. (7) holds true. $\hspace{0.1in} \square$

{\bf Proof of Theorem 2.3.}
``$(ii) \Rightarrow (i)$'' is evident. By the  monotonicity of coherence under
all SIOs, it is clear that ``$(i)\Rightarrow (iii)$''. We need only to  prove ``$(iii) \Rightarrow (ii)$''.


``$(iii)\Rightarrow (ii)$'' Suppose $C_{Re}(\rho)=C_{Re}(\sigma)$. Without loss of generality, we may also assume $\rho_{11}\geq \rho_{22}\geq \cdots\geq \rho_{dd}$ and  $\sigma_{11}\geq \sigma_{22}\geq \cdots\geq \sigma_{dd}$. The main result of \cite{Xiao} shows  $C_{l_1}(\rho)=C_{l_1}(\sigma)$ if $C_{Re}(\rho)=C_{Re}(\sigma)$.
From Theorem 2.1, it follows that $\sigma=\sum _{l}K_{l}\rho K_{l}^{\dag}$, $K_{l}=\sqrt{\delta_{l}}\sum_{i=1}^ d e^{i\theta_i^{(l)}}|f_{l}(i)\rangle \langle i|$. A direct computation shows
$$\Phi(\rho)_{mm}=\sum_{i}\sum_{l,f_{l}(i)=m } \delta_{l}\rho_{ii}.$$ Let $d_{mn}=\sum_{l,f_{l}(n)=m } \delta_{l}$. The the matrix $D$ with entries $d_{mn} (1\leq m,n\leq d)$ satisfies $$D(\rho_{11}, \rho_{22}, \cdots,\rho_{dd})^t=(\sigma_{11},\sigma_{22}, \cdots, \sigma_{dd},)^t.$$ Noting that $\sum_n d_{mn}=\sum_m d_{mn}=\sum_{l} \delta_{l}=1$, we know that $D$ is a bistochastic matrix.
This implies $(\sigma_{11},\sigma_{22}, \cdots, \sigma_{dd},)^t$ is majorized by $(\rho_{11}, \rho_{22}, \cdots,\rho_{dd})^t$ \cite{Bhatia}.
Recall that for two probability vectors $x=(x_1,\cdots , x_d)^t$ and $y=(y_1,\cdots , y_d)^t$, $x$ is majorized by $y$ if for each
$k$ in the range of $1, \cdots, d$, $$\sum_{i=1}^k x_i^\downarrow\leq
\sum_{i=1}^k y_i^\downarrow$$ with equality holding when $k= d$,
where the $x_i^\downarrow$ indicates that elements are to be taken
in descending order, so, for example, $x_1^\downarrow$ is the
largest element in $(x_1,\cdots , x_d)^t$. Therefore $\sum_{i=1}^k \sigma_{ii}\leq
\sum_{i=1}^k \rho_{ii}$ for each $1\leq k\leq d$.
In addition, combining $C_{Re}(\rho)=C_{Re}(\sigma)$ together with the Petz recovery map of \cite{Xiao}, we claim that $\Psi(\sigma)=\rho$ for some SIO $\Psi$. 
Thus there exists some bistochastic matrix $D'$ such that $$D'( \sigma_{11},\sigma_{22}, \cdots, \sigma_{dd})^t=(\rho_{11}, \rho_{22}, \cdots,\rho_{dd})^t.$$ This tells $(\rho_{11}, \rho_{22}, \cdots,\rho_{dd})^t$  is majorized by  $(\sigma_{11},\sigma_{22}, \cdots, \sigma_{dd},)^t$ and so $\sum_{i=1}^k \rho_{ii}\leq
\sum_{i=1}^k \sigma_{ii}$ for each $1\leq k\leq d$.
Thus $\rho_{ii}=\sigma_{ii}$ and $D=I$, i.e., $$d_{mn}=\sum_{l,f_{l}(n)=m } \delta_{l}=\left\{ \begin{array}{ll} 1   & m=n\\
                 0  & m\neq n.\end{array}\right.$$
This tells $f_{l}(i)=i$ for $1\leq i\leq d$. Combining this with $C_{l_1}(\rho)=C_{l_1}(\sigma)$, by a direct computation,
we have the following formulas: $$K_{l}=\sqrt{\delta_{l}}\sum_{i=1}^ d e^{i\theta_i^{(l)}}|i\rangle \langle i|,$$
$$\begin{array}{ll}&  \sum_{m\neq n} |\rho_{mn}|=\sum_{m\neq n}|\Phi(\rho)_{mn}|\\
 =&\sum_{m\neq n}|\sum_{l}\delta_{l}e^{i(\theta_m^{(l)}-\theta_n^{(l)})}\rho_{mn}|\\
\leq &\sum_{m\neq n}\sum_{l}\delta_{l}|e^{i(\theta_m^{(l)}-\theta_n^{(l)})}||\rho_{mn}|\\
= &\sum_{m\neq n}\sum_{l}\delta_{l}|\rho_{mn}|= \sum_{m\neq n} |\rho_{mn}|.\end{array}$$
Therefore $$|\sum_{\alpha}\delta_{l}e^{i(\theta_m^{(l)}-\theta_n^{(l)})}\rho_{mn}|=\sum_{l}\delta_{l}|e^{i(\theta_m^{(l)}-\theta_n^{(l)})}||\rho_{mn}|,$$
The triangle inequality of  complex numbers implies $$\theta_m^{(l)}-\theta_n^{(l)}=\theta_m^{(l')}-\theta_n^{(l')}.$$
So $K_{l}=\sqrt{\delta_{l}}\sum_{i=1}^ d e^{i\theta_i}|i\rangle \langle i|$. Let $U=\sum_{i=1}^ d e^{i\theta_i}|i\rangle \langle i|$. It is easy to check that $$\sigma=\sum_l K_l \rho K_l^\dag=U\rho U^\dag.\ \hspace{0.1in}\square $$

\vspace{0.1in}
{\bf Proof of Theorem 3.1.}
For the clarity of proof, let us  treat the the case  $d=4$, the general case can be treated similarly.
 The sufficiency is evident, we only check the necessity. Assume $\Phi(\rho)=\sigma$, we denote  $P_{\pi}\rho P_{\pi}^{\dag}$ and $P_{\pi}\sigma P_{\pi}^{\dag}$  by $\widetilde{\rho}$, $\widetilde{\sigma}$, and $\widetilde{\Phi}$ is the strictly incoherent operation with Kraus operators $P_{\pi}K_{\alpha}P_{\pi}^{\dag}$. Then $\widetilde{\Phi}(\widetilde{\rho})=\widetilde{\sigma}$ and $C_{l_1}(\widetilde{\rho})=C_{l_1}(\widetilde{\sigma})$.
By the properties of $\widetilde{\Phi}$, a direct computation shows the following formulas hold true:
$$\small{\begin{array}{ll}
&|\widetilde{\rho}_{12}|+|\widetilde{\rho}_{34}| =|\widetilde{\Phi}(\widetilde{\rho})_{12}|+|\widetilde{\Phi}(\widetilde{\rho})_{34}|\\
=& |\sum_{l_{f_{l}(1)=1,f_{l}(2)=2}}d_{l,1}\overline{d_{l,2}}\widetilde{\rho}_{12}+
\sum_{l_{f_{l}(2)=1,f_{l}(1)=2}}d_{l,2}\overline{d_{l,1}}\widetilde{\rho}_{21}\\
&+\sum_{l_{f_{l}(3)=1,f_{l}(4)=2}}d_{l,3}\overline{d_{l,4}}\widetilde{\rho}_{34}+
\sum_{l_{f_{l}(4)=1,f_{l}(3)=2}}d_{l,4}\overline{d_{l,3}}\widetilde{\rho}_{43}|\\
&+|\sum_{l_{f_{l}(1)=3,f_{l}(2)=4}}d_{l,1}\overline{d_{l,2}}\widetilde{\rho}_{12}+
\sum_{l_{f_{l}(2)=3,f_{l}(1)=4}}d_{l,2}\overline{d_{l,1}}\widetilde{\rho}_{21}\\
&+\sum_{l_{f_{l}(3)=3,f_{l}(4)=4}}d_{l,3}\overline{d_{l,4}}\widetilde{\rho}_{34}
+\sum_{l_{f_{l}(4)=3,f_{l}(3)=4}}d_{l,4}\overline{d_{l,3}}\widetilde{\rho}_{43}|\\
\leq &\sum_{l_{f_{l}(\{1,2\})=\{1,2\}\text{ or } \{3,4\}}}|d_{l,2}||d_{l,1}||\widetilde{\rho}_{12}|\\
&+\sum_{l_{f_{l}(\{3,4\})=\{1,2\}\text{ or } \{3,4\}}}|d_{l,3}||d_{l,4}||\widetilde{\rho}_{34}|\\
\leq&\sqrt{\sum_{l_{f_{l}(\{1,2\})=\{1,2\}\text{ or } \{3,4\}}}|d_{l,2}|^2}\times\\
&\sqrt{\sum_{l_{f_{l}(\{1,2\})=\{1,2\}}\text{ or } \{3,4\}}|d_{l,1}|^2}|\widetilde{\rho}_{12}|+\\
& \sqrt{\sum_{l_{f_{l}(\{3,4\})=\{1,2\}\text{ or } \{3,4\}}}|d_{l,3}|^2}\times\\
&\sqrt{\sum_{l_{f_{l}(\{3,4\})=\{1,2\}\text{ or } \{3,4\}}}|d_{l,4}|^2}|\widetilde{\rho}_{34}| \\ \leq & |\widetilde{\rho}_{12}|+|\widetilde{\rho}_{34}|.\end{array}}$$
Note that $\sum_l |d_{l,i}|^2=1 (i=1,2,3,4)$, so $f_{l}(\{1,2\})=\{1,2\}\text{ or } \{3,4\}$,  and $|d_{l,1}|=|d_{l,2}|$, $|d_{l,3}|=|d_{l,4}|$ for each $l$. This implies each $P_{\pi}K_{l}P_{\pi}^{\dag}$ has desired structure.
For arbitrary $m$, $$ \begin{array}{l}
\Big|\sum_{l_{f_{l}(\{2i-1,2i\})=\{2m-1,2m\}}}\delta_{l i}^2e^{i(\theta_{l i1}-\theta_{l i2})}\rho_{2i-1\ 2i}\Big|\\
=\sum_{l_{f_{l}(\{2i-1,2i\})=\{2m-1,2m\}}}\delta_{l i}^2|\rho_{2i-1\ 2i}|\end{array}$$ can be obtained directly from the above equalities.$\hspace{0.1in}\square$

{\bf Proof of Theorem 3.2.} By \cite{Xiao}, $C_{l_1}(\rho)=C_{l_1}(\sigma)$ if $C_{Re}(\rho)=C_{Re}(\sigma)$. Thus each $K_l$ has the same form as that of Theorem 3.1.
Write $\widetilde{\sigma}=\sum \lambda_i'\sigma_i$, here $$\lambda_i'=Tr(\sum _{l_{f_{l}(\{m,m+1\})=\{i,i+1\}}}\lambda_m{\delta}_{l m}^2U_{l m}\rho_m U_{l m}^{\dag}) ,$$ $$\sigma_i=\frac{1}{\lambda_{i}'}\sum _{l_{f_{l}(\{m,m+1\})=\{i,i+1\}}}{\delta}_{l m}^2U_{l m}\lambda_m\rho_m U_{l m}^{\dag} .$$ Note that $\tilde{\sigma}$ is a subspace independent state, by the additivity of relative entropy of coherence for subspace independent states \cite{Yuzhang} and the action of $\Phi$ on $\rho$,
$$\begin{array}{ll}
&C_{Re}(\sigma)=\sum_i \lambda_i'C_{Re}(\sigma_i)\\
\leq & \sum_i \lambda_i'\sum_{l_{f_{l}(\{m,d+1-m\})=\{i,d+1-i\}}} {\delta}_{l m}^2 \lambda_m C_{Re}(\rho_m)\\
=& \sum _m (\sum_{i,l_{f_{l}(\{m,d+1-m\})=\{i,d+1-i\}} }\lambda_i'{\delta}_{l m}^2 )\lambda_m C_{Re}(\rho_m)\\
=&\sum_{m} \lambda_m C_{Re}(\rho_m)=C_{Re}(\rho).\end{array}$$
By the equality for convexity of relative
entropy \cite{Jencova}, for given $m$,  $U_{l i}\rho_i U_{l i}^{\dag}=U_{s j}\rho_j U_{s j}^{\dag}$, here $f_{l}(\{i,i+1\})=f_{s}(\{j,j+1\})=\{m,m+1\}$.

The sufficiency is easy to see from the above inequality on $C_{Re}(\sigma)$.$\hspace{0.1in}\square$

{\bf Proof of Theorem 3.3.}  For the clarity, we only treat the the case $d = 5$, the general case can be done  similarly.
 Using similar argument as Theorem 3.1, we have either $f_{\alpha}(\{1,2\})=\{1,2\}\text{ or } \{3,4\}$, $f_{\alpha}(\{3,4\})=\{1,2\}\text{ or } \{3,4\}$ for each $\alpha$. Hence $$|d_{\alpha,1}|=|d_{\alpha,2}|\neq 0, |d_{\alpha,3}|=|d_{\alpha,4}|\neq 0$$ or $$f_{\alpha}(5)\neq 5, |d_{\alpha,1}|=|d_{\alpha,2}=|d_{\alpha,3}|=|d_{\alpha,4}|= 0.$$  One can finish the proof.$\hspace{0.1in}\square$

Using the similarly argument as  Theorem 3.2, Theorem 3.4 can be proved. $\hspace{0.1in}\square$

\end{document}